\documentclass[aps,prl,twocolumn,10pt,superscriptaddress,showpacs]{revtex4-1} 
\def\Title { Super Bloch oscillations with modulated interaction}

\usepackage{grffile}		
\usepackage[utf8x]{inputenc}
\usepackage[pdftex,colorlinks]{hyperref}	
\usepackage{amsmath,amsfonts,amssymb,bm}
\usepackage[pdftex]{graphicx}

\DeclareMathOperator{\IPN}{IPN}

\newcommand{\DepFisMat}{GISC, Departamento de F\'isica de Materiales, Universidad Complutense, E-28040 Madrid, Spain}

\newcommand\ie{\textit{i.e.}}
\newcommand\eg{\textit{e.g.}}

\begin{document}

\title{\Title}
\author{Elena Díaz}
\affiliation{\DepFisMat}
\author{Alberto García Mena}
\affiliation{\DepFisMat} 
\author{Kunihiko Asakura}
\affiliation{\DepFisMat}
\affiliation{Yonago National College of Technology, 683-8502, Yonago, Japan}
\author{Christopher Gaul}
\affiliation{\DepFisMat}
\affiliation{CEI Campus Moncloa, UCM-UPM, Madrid, Spain}
\date{\today} 
\begin{abstract}
We study super Bloch oscillations of ultracold atoms in a shaken lattice potential, subjected to a harmonically 
modulated mean-field interaction.
Usually, any interaction leads to the decay of the wave packet and its super Bloch oscillation.
Here, we use the phases of interaction and shaking with respect to the free Bloch oscillation as control parameters.
We find two types of long-living cases:
(i)  suppression of the immediate broadening of the wave packet, and
(ii) dynamical stability of all degrees of freedom.
The latter relies on the rather robust symmetry argument of cyclic time
[Gaul \emph{et al.}, Phys.\ Rev.\ A \textbf{84}, 053627 (2011)].
\end{abstract}


\maketitle

Experiments with ultracold atoms in optical lattices are an ideal testing ground for many problems of con\-densed-matter 
physics \cite{Lewenstein2007,Yukalov2009}.
In addition to good measurement access, these systems offer flexible manipulation of the system parameters, 
which opens the way to new perspectives and effects.
One of the most prominent examples is the observation of Bloch oscillations (BOs) of cold atomic gases \cite{BenDahan1996} 
and of Bose-Einstein condensates (BECs) \cite{Gustavsson2008} in optical lattices, subjected to an external force $F$.
The semiclassical explanation for BOs is the following:
the quasimomentum $\hbar k$ increases linearly with time, but because of the dispersion relation
of a tight-binding model with hopping amplitude $J$ and lattice period $a$, $E(k) = 2J[1-\cos(k a)]$, the group velocity is a sinusoidal function of time.
The particle does not follow the potential gradient but stays localized and performs an oscillatory motion.

By optical means or by magnetic levitation, the lattice potential can be shaken with frequency $\omega$.
This results in a renormalization of the hopping amplitude, which can even be suppressed, the so-called {dynamic localization} \cite{Dunlap1986,Holthaus1996}. 
Semiclassically, a harmonic shaking $F(t)=\Delta F \sin(\omega t)$ causes the quasimomentum $\hbar k = \int {\rm d}t' F(t')$ to 
oscillate rapidly and to explore $k$-space regions with renormalized or even negative effective mass $m_{\rm eff}(k) 
 \propto 1/\cos(k a)$.
Time-averaging $\cos(k a)$ leads to the renormalization of the hopping
$J_{\rm eff}/J = {\rm J}_0(\Delta F/\omega)$, where ${\rm J}_0$ 
is the Bessel function of the first kind.
Thus, depending on the strength and the frequency of the shaking, the effective hopping $J_{\rm eff}$ can be 
suppressed or even negative, freezing or inverting the center-of-mass motion.

\begin{figure}[b]
 \includegraphics[width=\linewidth]{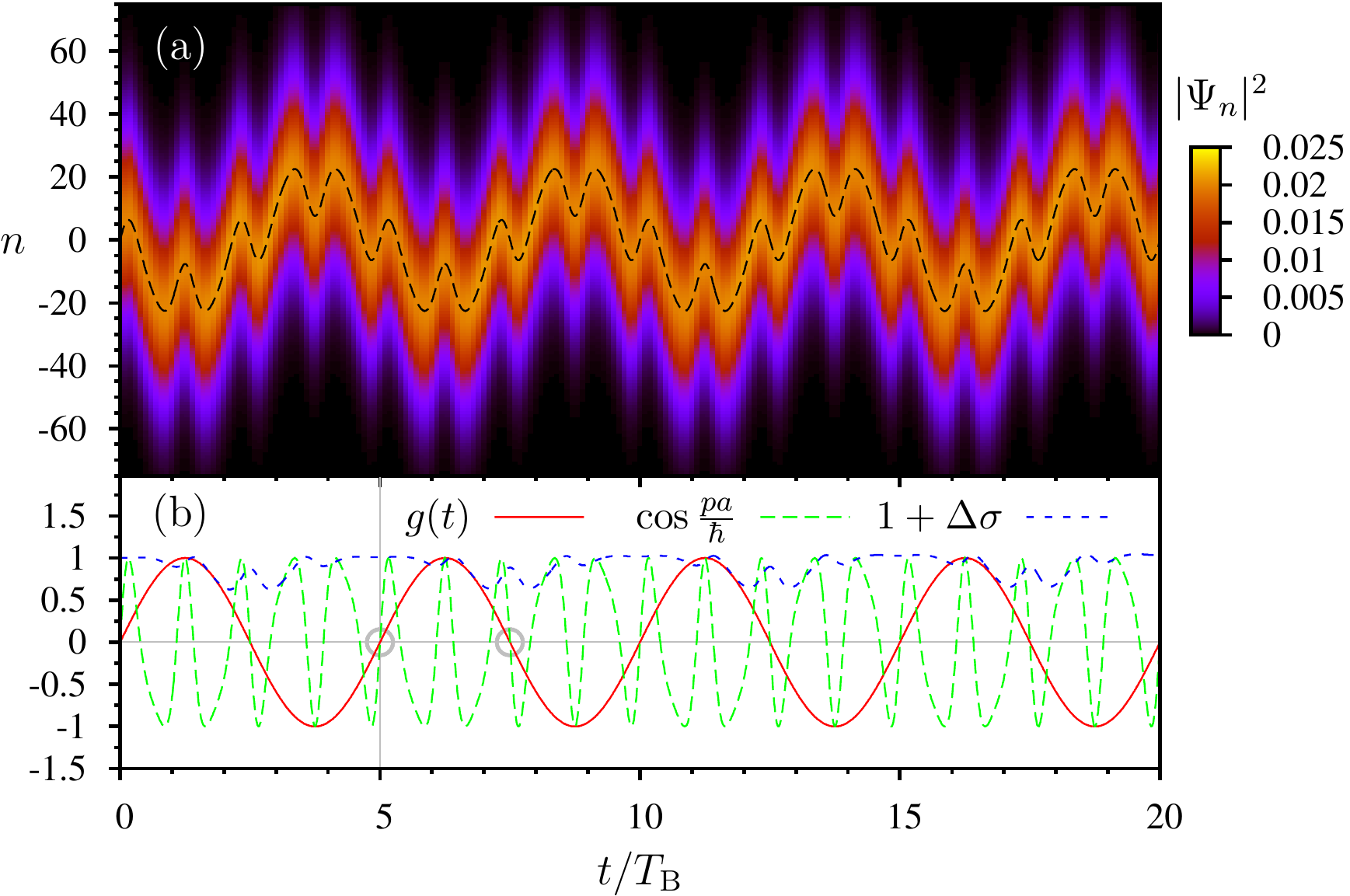}
 \caption{(Color online) Stable interacting SBO.
 The initial wave function has Gaussian shape with width $\sigma_0 = 20 a$ and quasimomentum $p(0)=\hbar \pi/2a$.
 Force \eqref{eqShaking} with $F_0=0.2 J/a$, $\Delta F=0.6 F_0$, $l/\nu=4/5$, 
 and $\phi_F=0$;
 interaction \eqref{eqInteraction} with $g_0=1$, $\phi=0$.
 (a)~Real-space density $|\Psi_n(t)|^2$ from the integration of Eq.~\eqref{eqTightBinding}.
 (b)~Time evolution of some magnitudes of interest: the interaction $g(t)$, the inverse mass term $\sim \cos(pa)$ and the variation of the real-space width $\Delta\sigma(t)=\sigma(t)-\sigma(0)$.
 The gray circles mark the points of symmetry used in the \it{cyclic-time approach}.
 }\label{figStableSBO}
\end{figure}

The modulation of the force around a finite mean value leads to a superposition of the BO
with a slow oscillation of large real-space amplitude, similar to the one shown in Fig.~\ref{figStableSBO}.
This phenomenon is known as quasi BO \cite{Wan2004} 
or super BO (SBO) \cite{Haller2010}.
It can be explained with a semiclassical reasoning, too:
the quasimomentum performs small oscillations around its linearly increasing mean value.
Then, during one Bloch cycle, it spends more time in $k$-space regions with, say, positive 
mass than in regions with negative mass, which
results in a drift of the wave packet in the direction of the force.
As the relative phase between BO and shaking changes, the time-averaged mass becomes negative and the drift gets reversed, which results in an SBO at the beating frequency.

SBOs have been mostly studied in the linear, noninteracting case \cite{Thommen2002,Arlinghaus2011}.
But ultracold bosons open more interesting possibilities.
Feshbach resonances can be used to arbitrarily change the $s$-wave scattering length \cite{
Chin2010},
even time-dependently \cite{Donley2001,Pollack2010,Vidanovic2011}.
%
Generically, the interaction leads to dephasing and decay of the wave packet.
However, the interplay of modulated interactions and BOs has already been investigated, and an 
infinite family of (harmonic) modulations that lead to a periodic time evolution of the wave packet 
has been found \cite{Gaul2009,Diaz2010,Gaul2011_bloch_long}.
Here, we consider SBOs in the presence of a modulated $s$-wave scattering length.
This problem is more complex because already the linear problem contains two frequencies and two phases from  
the BO and the shaking, respectively.
In the remainder of this Brief Report, we tackle the problem with a full integration of the discrete Gross-Pitaevskii equation and 
a cyclic-time argument similar to that of Ref.\ \cite{Gaul2011_bloch_long}, for one particular phase of the shaking.
Afterwards we study frequencies and phases not covered by the cyclic-time argument by means of \emph{collective coordinates} of a Gaussian wave packet and 
\emph{linear stability analysis} of the infinite wave packet.

\paragraph*{Model.} 
Our starting point is the mean-field tight-binding equation of motion as in Refs.\ \cite{Gaul2009,Diaz2010,Gaul2011_bloch_long}, but now the tilt $F$ may be time dependent:
\begin{equation}
\label{eqTightBinding}
i \hbar \dot\Psi_n = -J(\Psi_{n+1}+\Psi_{n-1}) + F(t) a n\Psi_{n} + g(t)|\Psi_{n}|^2\Psi_{n} .
\end{equation}
The wave function $\Psi_n$ is normalized to one, implying that the time-dependent interaction parameter $g(t)$ contains the total particle number.
We choose the force as 
\begin{equation}\label{eqShaking}
 F(t) = F_0 + \Delta F \cos(\omega t + \phi_F) .
\end{equation}
The mean value $F_0$ defines the Bloch frequency $\omega_{\rm B} = F_0a/\hbar$.
This and the frequency of the modulation $\omega$ are the two frequencies of interest.
For concreteness, we choose a fixed frequency ratio $\omega/\omega_{\rm B} = l/\nu = 4/5$ for the rest of this Brief Report. 
Thus, the super Bloch period $T_{\rm SBO}$ is five Bloch periods.
We now search for suitable modulations of $g(t)$ that allow a periodic time evolution of \eqref{eqTightBinding}.
A constant interaction parameter leads sooner or later to a decay of the SBO \cite{Haller2010}, 
but harmonic modulations around zero may counteract this effect \cite{Gaul2009}.
We choose the simplest commensurate modulation 
\begin{equation}\label{eqInteraction}
 g(t) = g_0 \sin(\omega_{\rm B} t/\nu + \phi) .
\end{equation}
Throughout the paper and we vary only the phase $\phi$.
Figure \ref{figStableSBO}(a) shows the time evolution of the wave function for $\phi_F=\phi=0$, a particular choice of 
parameters for which SBOs turn out to be stable.

\begin{figure}[bt]
\includegraphics[width=0.9\linewidth
]{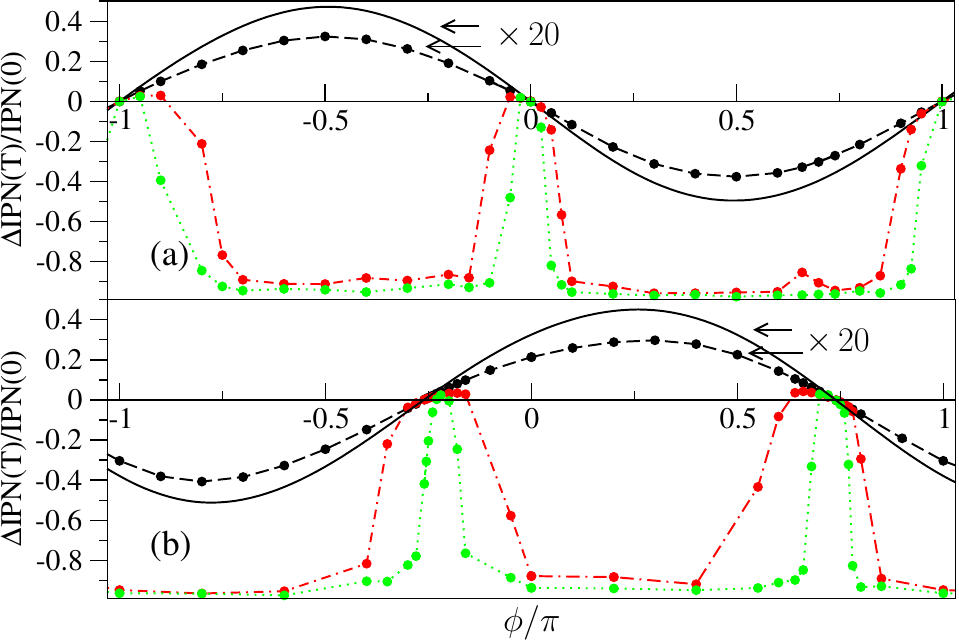}
 \caption{(Color online) 
Relative change of the momentum $\IPN$ as a function of the phase $\phi$ for (a) $\phi_F=0$ and (b)~$\phi_F=-\pi/2$. Discs (with lines as guide to the eye): results from integration of Eq.\ \eqref{eqTightBinding} after integration times $T=T_{\rm SBO}$, $15T_{\rm SBO}$, and $30 T_{\rm SBO}$ (from top to bottom). Solid line: Result from the collective coordinates theory \eqref{eqCCstable}. Regardless of a possible initial increase, the IPN always goes down in the long run, except for those points where $\Delta\IPN = 0$.
}\label{figBroadening}
\end{figure}

In general the SBO decays with time. 
This decay is most conveniently observed via the broadening of the wave-function momentum distribution,
observed via the inverse participation number (IPN) of Fourier modes 
\begin{equation}
 \IPN = \sum_k |\Psi_k|^4 . \label{eqParticipation}
\end{equation}
A decrease of this number means momentum broadening and decay of the wave packet.
Figure \ref{figBroadening} shows the variation $\Delta \IPN(T)=\IPN(T)-\IPN(0)$ after a certain time $T$ for $\phi_F=0,-\pi/2$ 
as a function of the interaction phase $\phi$.
For most of the phases, the $\IPN$ goes down, meaning that the wave packet broadens.
Figure \ref{figBroadening}(a) shows that for $\phi_F=0$ the phase $\phi=0$ from Fig.\ \ref{figStableSBO} is one of the two stable points, where the $\IPN$ stays constant.
Similarly, in the case $\phi_F=-\pi/2$ of Fig.\ \ref{figBroadening}(b), two ``stable'' points arise as well. However, their stability behavior 
is different and will be explained in what follows.


\paragraph*{Cyclic-time formalism for SBOs.}
We express the discrete wave function $\Psi_n$ of \eqref{eqTightBinding} in terms of a continuous wave function $A(z,t)$:
\begin{equation}\label{eqSmoothEnv}
\Psi_n(t) = A(n a - x(t),t) e^{i p(t) n a/\hbar + i \phi(t)} \ .
\end{equation}
With the 
semiclassical equations
$ \dot p = - F(t)$,
$ \dot x = v_{\rm g} = 2 J a \sin(pa/\hbar)/\hbar$,
and $\hbar \dot \phi = 2 J \cos(pa/\hbar)$, the envelope function obeys the nonlinear Schrödinger equation
\begin{align}\label{eqNLSE}
 i \hbar \dot A = - J \cos(p a/\hbar) \, A^{\prime\prime} + g(t) |A|^2 A \ ,
\end{align}
where higher derivatives of $A$ have been neglected.
The factor in front of the second derivative takes the role of the mass term $J \cos(p a/\hbar) = {\hbar^2}/{2m(t)}$.
Taking into account the commensurability of the frequencies $\omega$ and $\omega_{\rm B}$, 
we express the time dependence in terms of the slower time $\tau = \omega_{\rm B} t/\nu$,
which increases by $2\pi$ during one super Bloch period $T_{\rm SBO} = \nu T_{\rm B}$.
After integrating the force \eqref{eqShaking}, we find the inverse mass term as
\begin{equation}\label{eqcosp}
\cos(pa/\hbar) = \sin\left\lbrace \nu \tau + \textstyle{\frac{a\, \Delta F}{\hbar\omega}} \left[\sin(l \tau+\phi_F) - \sin \phi_F \right]\right\rbrace,
\end{equation}
where we have chosen the initial condition such that $a\, p(\tau=0) = \pi\hbar/2$.
This can always be achieved by choosing the origin of time.

We know from previous works \cite{Gaul2009,Diaz2010,Gaul2011_bloch_long} that BOs can exist in the presence of suitably 
tuned AC interactions.
Now, we construct a similar cyclic-time formalism for SBOs.
The basic idea is to separate all terms of Eq.\ \eqref{eqNLSE} into the form $\dot \eta f(\eta)$, where $\eta$ is a 
harmonic function of $\tau$ and $f$ an arbitrary function.
Then $\dot \eta$ is eliminated and Eq.\ \eqref{eqNLSE} is solved as a function of $\eta$ only.
As $\eta$ is a periodic function of $t$, the time evolution of $A$ is periodic too, and we have indeed found a case of 
stable SBOs in the presence of a non-zero interaction $g(t)$.

The inverse mass term \eqref{eqcosp}
can only be brought into the desired form $\dot \eta f(\eta)$ if $\phi_F = 0$ [or any phase shift that allows for a common node 
of $\sin \nu \tau$ and $\sin l \tau$]. Then,
\begin{equation}\label{eqCyclic}
\cos(p a/\hbar) 
= \sin(\tau) \mathcal{F}(\cos\tau) ,
\end{equation}
which is the most general $2\pi$-periodic function that respects the odd symmetry with respect to $\tau = 0 \mod \pi$.
This determines $\eta = \cos \tau$.
Thus, the left hand side of \eqref{eqNLSE} is $\dot A = \dot \eta \partial_\eta A$.
Finally, we may choose any modulated interaction of the form $g=\dot\eta \tilde g(\eta)$. 
In particular, $g(t) = g_0 \sin(F_0 t /\nu)$, \ie, Eq.\ \eqref{eqInteraction} with $\phi=0$ 
fulfills the cyclic-time condition.
The symmetries involved in this argument can be observed 
in the lower panel of Fig.\ \ref{figStableSBO}:
$g(t)$ and $\cos\bigl(p(t)a/\hbar\bigr)$ share the nodes of $\dot \eta$, marked with gray circles, and are odd with respect to these points.
All physical quantities, \eg\ the width variation $\Delta \sigma$, are functions of $\eta$ only and are thus even with respect to the mentioned points.

So far, we have explained the stability observed in Fig.~\ref{figStableSBO} and the stable points of Fig.\ \ref{figBroadening}(a).
It turns out that the phase $\phi_F=-\pi/2$ of shaking considered in Fig.\ \ref{figBroadening}(b) 
cannot be written in the form required by the above cyclic-time argument, because the inverse
mass term $\cos a p/\hbar$ does not exhibit points with odd symmetry.
Nevertheless, there are two particular phase shifts, $\phi \approx -0.26\pi$ and $\phi \approx 0.74\pi$, 
where the broadening of the wave packet is suppressed.
One can understand this using a
\emph{collective-coordinates theory}.
The idea is to reduce the complexity of the problem by parametrizing 
the wave function by only two coordinates, its position $x$ and its width $\sqrt{w}$ \cite{Trombettoni2001,Gaul2009}:
\begin{align}\label{psiA.eq}
\Psi_n(t) = \frac{1}{\sqrt[4]{w}}\, {\mathcal{A}} 
\left(\frac{n a - x}{\sqrt{w}}\right) 
e^{i p n a / \hbar +i b (n a-x)^2/\hbar
} .
\end{align}
$\mathcal{A}(u)$ is a continuous normalized Gaussian wave function with variance one.
The equations of motion for the collective coordinates, $x$ and $w$, and their conjugate momenta, $p$ and $b$, are
\begin{subequations}
\begin{align}
\dot p &= -F(t)\, ,    \label{pdot} \\
\frac{\hbar}{J} \, \frac{\dot x}{a} &= 2 {a}\sin(p a/\hbar) \left[1- \frac{1/4+4b^2w^2 /\hbar^{2}}{2w /a^{2}}\right] , \label{xdot}\\
\frac{a^2}{J}\, \dot b &= \frac{1-16w^2b^2/\hbar^2}{4w^2/a^4}\cos(p a/\hbar) +  \frac{a^3 g(t)}{8\sqrt{\pi} J w^{3/2}}\, , \label{bdot} \\    
\frac{\hbar \dot w}{J a^2}  &= 8 \frac{w b}{\hbar} \cos(p a/\hbar)\ .   \label{wdot}  
\end{align}
\end{subequations}
A necessary condition for stability within the collective-coordinates approximation is that $w(t)$ returns to 
its initial value after a full super Bloch period $T_{\rm SBO}$.
In this regard, we can compute analytical results in the limit of a wide wave packet $w(t) \approx \sigma_0^2 \gg 1$. 
Then, Eqs.\ \eqref{eqInteraction} and \eqref{bdot} give
$b(t) \approx - {a \nu g_0}/[{8\sqrt{\pi} \sigma_0^3 \omega_{\rm B}}] \cos( \omega_{\rm B} t/\nu +\phi)$.
With \eqref{wdot}, and assuming that $|l-\nu|=1$, this yields
\begin{equation}\label{eqCCstable}
 \int_{T_{\rm SBO}} {\rm d}t \frac{\dot w}{w} 
 \approx (l-\nu) \frac{\sqrt{\pi} a^3 J g_0}{\sigma_0^3 (\hbar\omega_{\rm B}/\nu)^2} 
    {\rm J}_1\bigl(\textstyle{\frac{\delta F \nu}{l}}\bigr)
    \sin\left(\phi - \phi_0 \right) ,
\end{equation}
with $\phi_0 = (l-\nu) \left[\phi_F + a \Delta F  \sin(\phi_F) /\hbar\omega \right]$.
Depending on the phase $\phi$ of the interaction, the wave packet contracts or spreads after a full super Bloch cycle $T_{\rm SBO}$.
This is also reflected in the $\IPN$ shown in Fig.\ \ref{figBroadening}.
The dynamics can be strictly periodic only for $\phi=\phi_0 \mod \pi$.
With the values $l=4$, $\nu=5$, $\phi_F=-\pi/2$, and 
$a\Delta F =0.75 \hbar \omega$, this yields $\phi_0=0.74\pi$, in agreement with Fig.\ \ref{figBroadening}(b).

\begin{figure}
 \includegraphics[width=\linewidth]{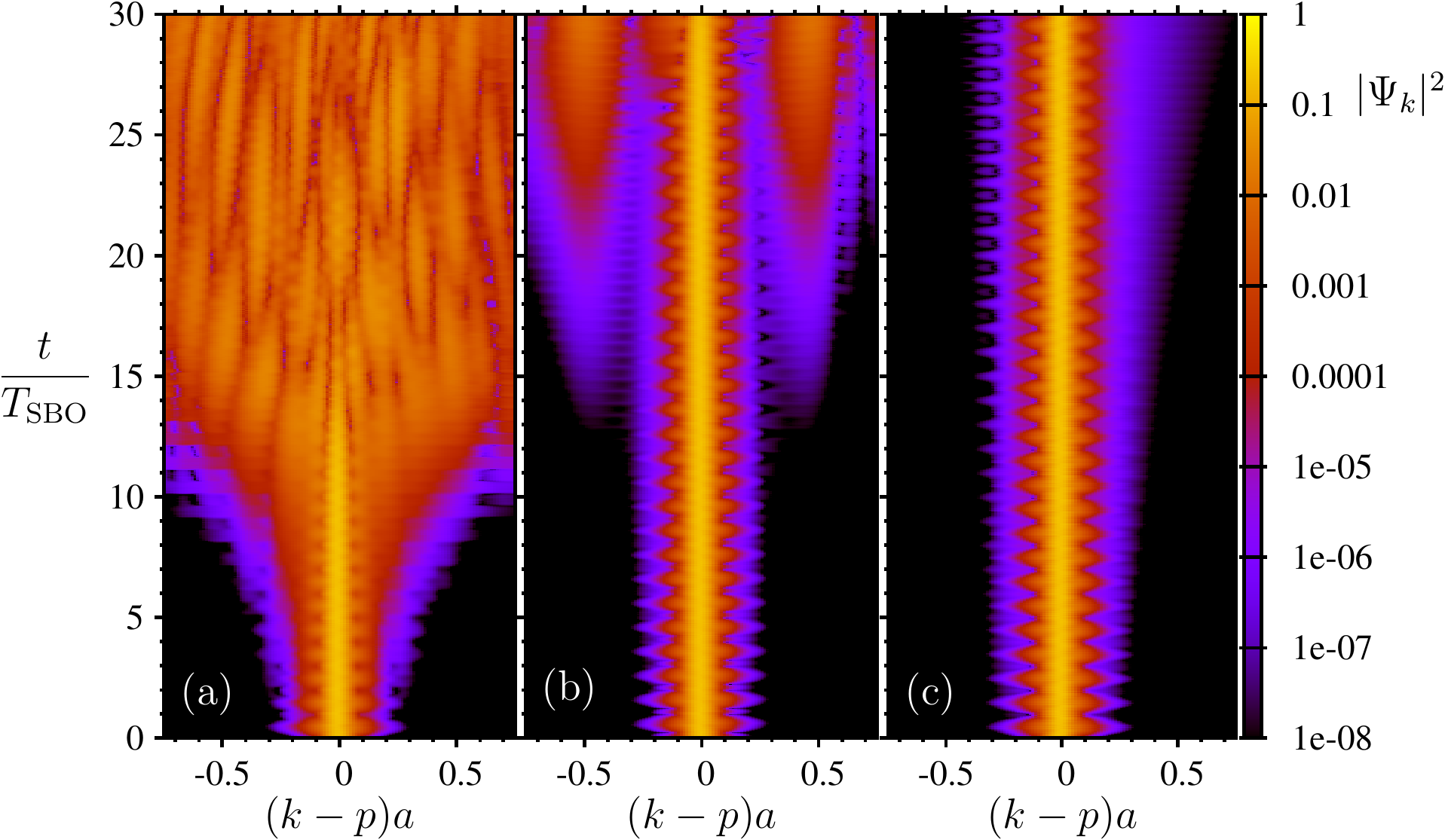}
\caption{(Color online) 
 Momentum-space portraits in the reference frame of the center of momentum $p$ for shaking \eqref{eqShaking} and interaction \eqref{eqInteraction},
 with $\sigma_0=20 a$, $F_0=0.2 J/a$, $\Delta F = 0.6 F_0$, 
 $g_0=1$.
 (a) Unstable sine shaking $\phi_F=-\pi/2$ with $\phi=0$;
 (b) sine shaking with phase $\phi=-0.26\pi$ adjusted according to Eq.~\eqref{eqCCstable};
 (c) stable case of Fig.\ \ref{figStableSBO}, $\phi_F=0$, $\phi=0$.
\label{figkspace}}
\end{figure}

In Fig.\ \ref{figkspace}, the momentum-space portrait of SBOs is shown for different phases of the shaking and of the interaction.
Panel (a) shows an unstable case of sine shaking with immediate broadening of the momentum distribution.
In panel (b) the phase of the interaction is adjusted as required by the collective-coordinates criterion of Eq.\ \eqref{eqCCstable}
and it shows the behavior presented by the two apparently ``stable'' points in panel (b) of Fig.\ \ref{figBroadening}.
Here, the immediate broadening is indeed suppressed; but on a longer time scale ($\approx 20 T_{\rm SBO} = 100 T_{\rm B}$), 
side peaks grow far from the central wave packet.
This is an indicator for a dynamical instability and does not happen in the stable case with cosine shaking 
(see Fig.\ \ref{figStableSBO}), whose $k$-space portrait is shown in Fig.\ \ref{figkspace}(c) for comparison. 
In this case, there is neither immediate broadening nor an instability. 
The extremely small residual broadening, is due to effects beyond the approximation made in Eq.~\eqref{eqNLSE}. 

\paragraph*{Linear stability analysis.}
\begin{figure}[tb]
 \includegraphics[width=\linewidth]{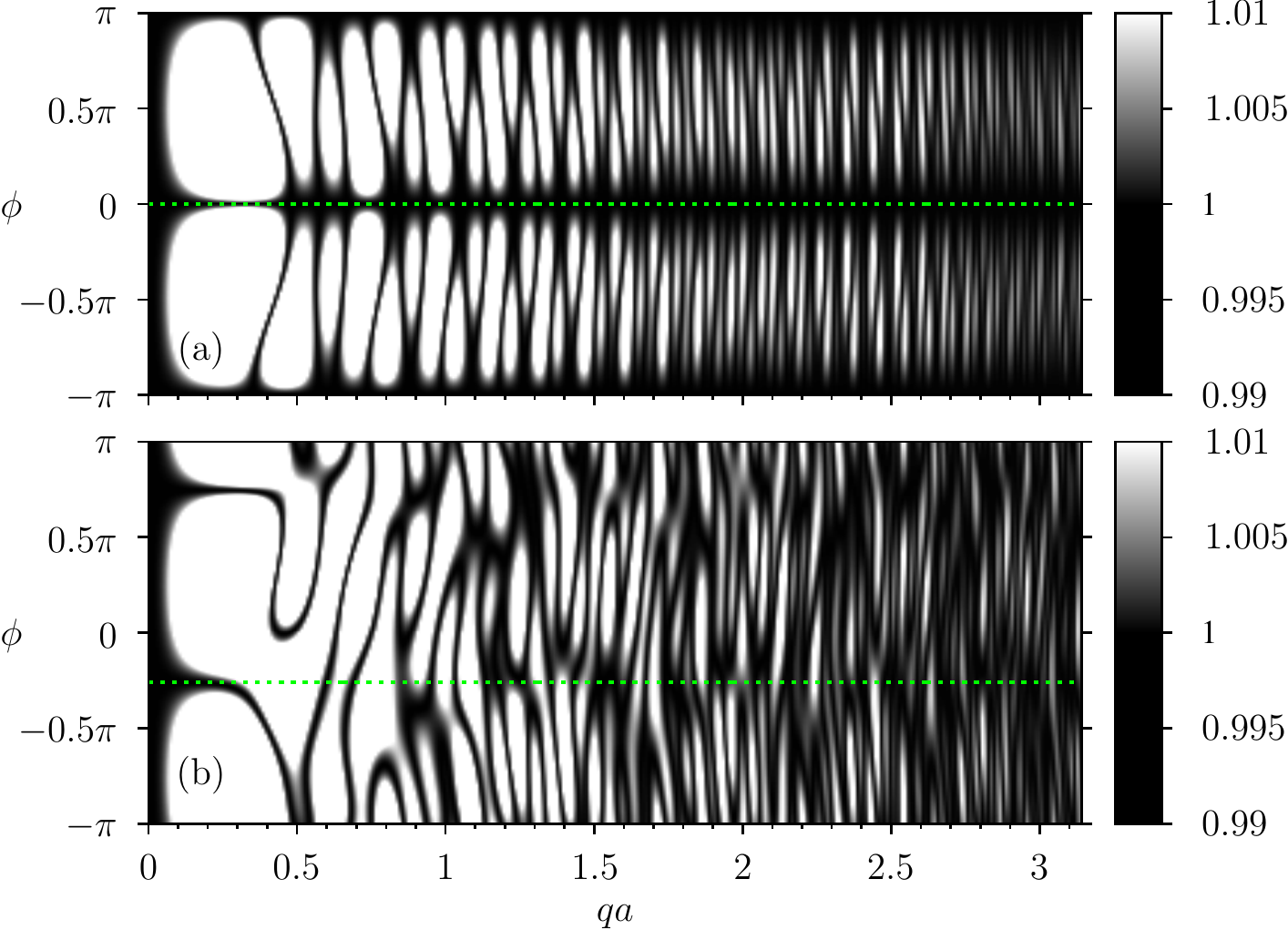}
\caption{(Color online) 
 Stability prediction $\Delta_q$ (black: stable, white: unstable) for fluctuation mode $q=k-p$ 
 as a function of the interaction phase $\phi$ for cosine shaking $\phi_F=0$ (a) and sine shaking $\phi_F=-\pi/2$ (b).
 The dashed green line indicates the phase $\phi_g$ taken in Figs.\ \ref{figkspace} (b) and (c), respectively.
\label{figStability}}
\end{figure}
Collective coordinates rely on a smooth Gaussian profile of the wave function and cannot describe features on short length scales.
We close this gap with a linear stability analysis of the infinitely extended Bloch oscillating wave packet.
The method is the same as in Ref.\ \cite{Gaul2009}, only that $\cos(Ft)$ is replaced with
$\cos(pa/\hbar)$ from Eq.\ \eqref{eqcosp} and the period with the super Bloch period.
For each plane-wave modulation $q$ of the wave function, the monodromy matrix is obtained by integration of the (linearized) equations of motion over one period with two different initial conditions.
The mode $q$ is stable if $\Delta_q$, which is half the trace of the monodromy matrix, is less or equal to one.
Conversely, it is unstable if $|\Delta_q|>1$ \cite{Teschl2012}.
Figure \ref{figStability} shows maps of stability for both cases of primary interest, cosine shaking 
($\phi_F=0$) and sine shaking ($\phi_F=-\pi/2$).
In the first case, for values $\phi = n \pi$, $n\in\mathbb{Z}$, all modes $q$ remain stable, in concordance 
with the cyclic-time argument and with Fig.\ \ref{figBroadening}(a).
In the second case, the stable phase of collective coordinates agrees with the stability for small values of $q$ only,
and for larger values instabilities occur.
Notably, the first region of instability is located around $q=0.5$. This is just the region where the instability 
occurs in the full integration, as shown in Fig.\ \ref{figkspace}(b).

\paragraph*{Conclusions.}
We have studied the stability of commensurate super Bloch oscillations under a time-dependent interaction parameter 
that is modulated with the super Bloch frequency.
The phase of the interaction can always be adjusted such that the direct broadening of the width is suppressed
[see Fig.\ \ref{figBroadening} and Eq.\ \eqref{eqCCstable}].
But there are more degrees of freedom, which may be subject to dynamical instability [Fig.\ \ref{figkspace}(b)].
These can be stabilized by fulfilling the cyclic-time argument [Fig.\ \ref{figkspace}(c)].

The dynamics of BOs and SBOs strongly depends on the relative phase evolution between neighboring sites \cite{Gustavsson2008,Haller2010}.
This means that interactions distort the density profile of the driven Bose-Einstein condensate and lead to the
destruction of such oscillations in general. Therefore, our proposal to tune the atomic interaction time dependently
is important to reduce this effect and to allow for the detection of such oscillatory dynamics of a condensate. 
The results of this work can be considered to improve the quality of SBOs in order to be used 
to engineer matter-wave transport over macroscopic distances in lattice potentials
in a more reliable way, which should also be relevant for atom interferometry~\cite{Cronin2009}.

\begin{acknowledgments} 
This work was supported by Ministerio de Economía y Competitividad (project MAT2010-17180).
Research of C.G.\ was supported by a PICATA postdoctoral fellowship
from the Moncloa Campus of International Excellence (UCM-UPM).
\end{acknowledgments}

\bibliography{references_sbo}

\end{document}